\documentclass[11pt]{article}
\begin{document}
\def\doublespaced{\baselineskip=\normalbaselineskip\multiply\baselineskip
 by 150\divide\baselineskip by 100}
\doublespaced
\def\lsim{~{\rlap{\lower 3.5pt\hbox{$\mathchar\sim$}}\raise 1pt\hbox{$<$}}\,}
\def\gsim{~{\rlap{\lower 3.5pt\hbox{$\mathchar\sim$}}\raise 1pt\hbox{$>$}}\,}
\def\thisday{~September~10, 1998 ~and~ hep-th/yymmnnn~~}
\def\thisday{~\today ~and~ hep-pth/yymmnnn~~}


\vbox to 3.5cm {
\vfill
}
\begin{center}{\Large{Fractional Integral and Derivative of the $1/r$ Potential }}\\
Ehab Malkawi  \\
{{Department of Physics,\\
United Arab Emirates University\\
 Al Ain, UAE}}\\
 emalkawi@uaeu.ac.ae
\end{center}
\vskip -10pt
\begin{center}
Abstract
\end{center}
We calculate the fractional integral and derivative of the potential $1/r$ for all values of the fractional order $-1< \alpha \leq 0$ and $\alpha\geq 0$. We show that the result has the same form for all values of $\alpha$. Applications can be implemented to discuss deformed
potential fields resulting from fractional mass or charge densities in gravity and electrostatics problems. The result can also be applied  to modify the inverse-square law gravity as predicted by new physics.


\vskip 6pt



\vskip 3pt

\newpage

\section{Introduction}

Fractional calculus deals with differentiation and integration to arbitrary real or complex orders.  Extensive mathematical discussion of fractional calculus can be found in Refs.~[1]-[4] and references therein.
The techniques of fractional calculus have been applied to wide range of fields, such as physics, engineering, chemistry, biology, economics, control theory, signal image processing, groundwater problems, and many others.

Physics applications of fractional calculus span a wide range of topics and problems (for a review see Refs.~[5]-[11] and references therein). Generalizing fractional calculus to several variables, multidimensional space, and  generalization of  fractional vector calculus has been reported [12]-[19].  Also, progress has been reported on generalization of Lagrangian and Hamiltonian systems [20]-[25].

In this work we simply consider the potential field $\Phi(r)=k/r$, where $k$ is constant describing the strength of the field and $r=\sqrt{x^2+y^2+z^2}$. The potential field emerges from an inverse-square law of gravity and Coulomb electric field. We calculate the fractional integral and derivative of $\Phi$ using Cartesian coordinates and for wide range of the fractional order $\alpha$. This calculation is important for applications to gravity and electrostatics problems. For example, one can implement
the techniques of fractional calculus to relate two known mass or charge distributions by a continuous
deformation as discussed in Refs.~[26, 27]. Thus, one can study these deformations (intermediate distributions) and their corresponding
intermediate gravitational or electrical potentials.

Modifications to the inverse-quare law gravity has been are argued in theories of large extra dimensions, broken supersymmetry at low energy, and string theories, while deviation has been tested in several experiments (see Refs. [28]-[30] and references therein).
In classical gravity, one can think of the inverse-square force as emerging from the specific potential field $\Phi(r)$. The two views are equivalent as they are related by the integer derivative operator, the gradient, $\textbf{F}=-\nabla \Phi$. A possible slight deviation from the inverse square force could be due to a slight modification to the integer derivative, in other words, a fractional derivative of the potential field will lead to modification to the inverse-square force. This motivates the need to consider the fractional derivative of the potential field $\Phi$. A different approach in modifying the inverse-square law is within fractional space [31] and where gravitational field is derived from a fractional mass distribution [32].

Finally the analytical derivation of the fractional derivative of $1/r$ in Cartesian coordinates could shed light on the corresponding connection with the fractional derivative in other coordinate systems, such as spherical coordinates [33].
In the next section we lay out the basic definitions of fractional calculus relevant to our study. In Section 3 we calculate the fractional integral and derivative of $\Phi(r)$. Finally, in Section 4 we provide some discussion.

\section{Fractional Calculus}

For mathematical properties of fractional derivatives and integrals one can consult Refs.[1]-[5] and the references therein.
In this section we
lay out the notation used in the next section as we consider the Riemann-Liouville and Caputo definitions of the fractional derivative. In this work both definitions give identical results.
Let $f(x,y,z)$ to be a real analytic function in a specific domain in the Euclidian space $R^3$; $f: R^3 \rightarrow R$.
The $x$-partial fractional integral or derivative of order $\alpha$ (keeping $y$ and $z$ constants) is written as $_aD_x^\alpha f(x,y,z)$, where $a$ is the
lower limit of $x$. Similarly the $y$- and $z$-partial fractional integral or derivatives of order $\alpha$ are written as $_bD_y^\alpha f(x,y,z)$ and $_cD_z^\alpha f(x,y,z)$, respectively. Note that $\alpha<0$ represents a fractional integral, while $\alpha>0$ represents a fractional derivative.
Since $f(x,y,z)$ is analytic then the partial fractional derivatives are assumed to commute, i.e., $\left[ _aD_x^\alpha,\, _bD_y^\alpha\right]=\left[ _aD_x^\alpha,\, _cD_z^\alpha\right]=\left[ _bD_y^\alpha,\, _cD_z^\alpha\right]=0$. \\
\textbf{Definition 2.1.}
The Cauchy's repeated integration
formula of the $n$th-order integration of the function $f(x,y,z)$ along $x$, keeping $y$ and $z$ constants, can be written as
\begin{eqnarray}
_aD_x^{-n} f(x,y,z)&=& \int_a^x dx_{n-1} \int_a^{x_{n-1}} dx_{n-2}\dots \int_a^{x_{1}} f(x_0,y,z)\,dx_0
 \nonumber \\
&=&\frac{1}{(n-1)!}\int_a^x \frac{f(u,y,z)\,du}{{(x-u)}^{1-n}}\,\, .
\end{eqnarray}
A similar formula for the\, $n$th-order integration of the function $f(x,y,z)$ along $y$, keeping $x$ and $z$ constants,
\begin{eqnarray}
_bD_y^{-n} f(x,y,z)&=v& \int_b^y dy_{n-1} \int_b^{y_{n-1}} dy_{n-2}\dots \int_b^{y_{1}} f(x,y_0,z)\,dy_0
 \nonumber \\
&=&\frac{1}{(n-1)!}\int_b^y \frac{f(x,u,z)\,du}{{(y-u)}^{1-n}}\,\, .
\end{eqnarray}
Similarly for the\, $n$th-order integration of the function $f(x,y,z)$ along $z$, keeping $x$ and $y$ constants,
\begin{eqnarray}
_cD_z^{-n} f(x,y,z)&=v& \int_c^z dz_{n-1} \int_b^{z_{n-1}} dz_{n-2}\dots \int_c^{z_{1}} f(x,,y,z_0)\,dz_0
 \nonumber \\
&=&\frac{1}{(n-1)!}\int_c^z \frac{f(x,y,u)\,du}{{(z-u)}^{1-n}}\,\, .
\end{eqnarray}

\textbf{Definition 2.2.} The fractional integration of order $\alpha<0$ and along $x$, keeping $y$ and $z$ constants,
is defined as
\begin{equation}
_aD_x^\alpha f(x,y,z)= \frac{1}{\Gamma(-\alpha)} \int_a^x \frac{f(u,y,z)\,du}{{(x-u)}^{1+\alpha}} \,\, .
\label{eq0}
\end{equation}
Similarly the fractional integration along $y$, keeping $x$ and $z$ constants, is
\begin{equation}
_bD_y^\alpha f(x,y,z)= \frac{1}{\Gamma(-\alpha)} \int_b^y \frac{f(x,u,z)\,du }{{(y-u)}^{1+\alpha}}\, \, .
\label{eq01}
\end{equation}
Similarly the fractional integration along $z$, keeping $x$ and $y$ constants, is
\begin{equation}
_cD_z^\alpha f(x,y,z)= \frac{1}{\Gamma(-\alpha)} \int_c^z \frac{f(x,y,u)\,du }{{(z-u)}^{1+\alpha}}\, \, .
\label{eq02}
\end{equation}
where $\Gamma(.)$ is the Gamma function.

\textbf{Definition 2.3.} The Riemann-Liouville partial fractional derivatives of the order $\alpha>0$, where $n-1<\alpha<n$ and $n\in N$, are defined as
\begin{eqnarray}
_aD_x^\alpha f(x,y,z) = \frac{\partial^n}{\partial x^n} \, {_aD_x^{\alpha-n}} f(x,y,z) =
\frac{1}{\Gamma(n-\alpha)} \frac{\partial^n}{\partial x^n} \int_a^x \frac{f(u,y,z)\, du}{{(x-u)}^{\alpha-n+1}}\,\, , \\
_bD_y^\alpha f(x,y,z) = \frac{\partial^n}{\partial y^n} \, {_bD_y^{\alpha-n}} f(x,y,z) =\frac{1}{\Gamma(n-\alpha)} \frac{\partial^n}{\partial y^n} \int_b^y \frac{f(x,u,z)\, du}{{(y-u)}^{\alpha-n+1}}\,\, , \\
_cD_z^\alpha f(x,y,z) = \frac{\partial^n}{\partial z^n}\, {_cD_z^{\alpha-n}} f(x,y,z) =\frac{1}{\Gamma(n-\alpha)} \frac{\partial^n}{\partial z^n} \int_c^z \frac{f(x,y,u)\, du}{{(z-u)}^{\alpha-n+1}}\,\, .
\label{derivative}
\end{eqnarray}

\textbf{Definition 2.4. }
The Caputo partial fractional derivatives of order $\alpha>0$, where $n-1<\alpha<n$ and $n\in N$, are defined as
\begin{eqnarray}
  ^{C}_aD_x^\alpha f(x,y,z)   =  \frac{1}{\Gamma(n-\alpha)} \int_a^x   \frac{ \frac{\partial^n}{\partial u^n} f(u,y,z)\, du}{{(x-u)}^{\alpha-n+1}} \,\, ,
  \label{cap1}\\
  ^{C}_bD_y^\alpha f(x,y,z)   =    \frac{1}{\Gamma(n-\alpha)} \int_b^y \frac{ \frac{\partial^n}{\partial u^n} f(x,u,z)\, du}{{(y-u)}^{\alpha-n+1}}\,\, ,\\
 ^{C}_cD_z^\alpha f(x,y,z)   =    \frac{1}{\Gamma(n-\alpha)} \int_c^z \frac{ \frac{\partial^n}{\partial u^n} f(x,y,u)\, du}{{(z-u)}^{\alpha-n+1}}\,\, .
\label{cap2}
\end{eqnarray}

The Riemann-Liouville and Caputo definitions of the fractional derivative are related [1]-[5]. In our work, we consider the lower limit $a=b=c=-\infty$ and
since all partial derivatives of $\Phi$ vanish at the lower limit, we conclude that Riemann-Liouville and Caputo definitions of the fractional derivatives are equivalent, giving rise to the same result.

\section {Fractional integral and derivative of $1/r$ potential}

 We consider the potential field $\Phi(r)=k/r$, where $k$ is constant describing the strength of the field and $r=\sqrt{x^2+y^2+z^2}$.
 In deriving the fractional integral and derivative of $\Phi(r)$ we choose the lower limit of the fractional integral to be $a=b=c=-\infty$. We will drop the constant $k$ in our derivation and can later be inserted with its correct dimensionality according to specific applications.

 \subsection{Fractional integral of $1/r$}

 We start by calculating the fractional integral along $z$ for $-1<\alpha\leq 0$. According to Eq.~(\ref{eq02}) 
\begin{eqnarray}
 _{-\infty}{D_z^\alpha} \Phi(x,y,z) &=&  \frac{1}{\Gamma(-\alpha)}
 \int_{-\infty}^z \frac{\Phi(x,y,u)\,du }{{(z-u)}^{\alpha}}\nonumber \\
  &=& \frac{1}{\Gamma(-\alpha)}\int_{-\infty}^z \frac{{(x^2+y^2+u^2)}^{-1/2}\,du }{{(z-u)}^{\alpha+1}}
 \, \, .
\end{eqnarray}
Write $\rho^2 =x^2+y^2$ and let $t=z-u$ we get
\begin{eqnarray}
_{-\infty}{D_z^\alpha} \Phi(x,y,z)=
\frac{1}{\Gamma(-\alpha)}\int_0^{+\infty} \frac{{(\rho^2+z^2-2zt+t^2)}^{-1/2}\,dt }{t^{\alpha+1}}
 \, \, .
\end{eqnarray}
Using the spherical coordinates, $r$ and $\theta$, where $r^2=\rho^2+z^2$ and $z=r\cos\theta$ we get
\begin{eqnarray}
_{-\infty}{D_z^\alpha} \Phi(x,y,z)=
\frac{1}{\Gamma(-\alpha)}\int_0^{+\infty} \frac{{(r^2-2rt\cos\theta+t^2)}^{-1/2}\,dt }{t^{\alpha+1}}
 \, \, .
\end{eqnarray}
Dividing the integration  into the two regions $0<t<r$ and $t>r$, we expand the integrand in terms of Legendre polynomials. Integrating over $t$
we get the final form
\begin{eqnarray}
_{-\infty}{D_z^\alpha} \Phi(x,y,z)= \frac{1}{\Gamma(-\alpha)\,{r^{\alpha+1}}}\sum_{n=0}^\infty \frac{2n+1}{n(n+1) -\alpha(\alpha+1)} P_n(\cos\theta)\, .
\label{integral}
\end{eqnarray}
The result is divergent for $\theta= 0$ (i.e., $x=y=0$ and $z>0$) and convergent everywhere else. 
The result agrees with Ref.~[27] for $-1<\alpha<0$. For $\alpha = 0$, we have $1/\Gamma(-\alpha)=-\alpha+O(\alpha^2)$ and it is easy to check that we retrieve the original field $1/r$, as all terms vanish except the first term in the series (n=0).

Due to the spherical symmetry of the potential, one can easily conclude that
\begin{eqnarray}
_{-\infty}{D_x^\alpha} \Phi(x,y,z)=
 \frac{1}{\Gamma(-\alpha)\, r^{\alpha+1}} \sum_{n=0}^\infty \frac{2n+1}{n(n+1)-\alpha(\alpha+1)} P_n(\sin\theta\cos\phi)\, ,\\
 _{-\infty}{D_y^\alpha} \Phi(x,y,z)=
 \frac{1}{\Gamma(-\alpha)\, r^{\alpha+1}} \sum_{n=0}^\infty \frac{2n+1}{n(n+1)-\alpha(\alpha+1)} P_n(\sin\theta\sin\phi)\, ,
 \label{final3}
 \end{eqnarray}
 where $\phi$ is the azimuthal angle in the spherical coordinates, $x=r\sin\theta \cos\phi$ and $y=r\sin\theta \sin\phi$.

\subsection{Fractional derivative of $1/r$}
In deriving the fractional derivative of $\Phi(r)$ we choose the lower limit to be $-\infty$. Since all partial derivatives of $\Phi$ vanish at the lower limit, we conclude that Riemann-Liouville and Caputo definitions are equivalent, giving rise to the same result.
We consider first $0<\alpha<1$, according to Eq.~(\ref{derivative})
\begin{eqnarray}
_{-\infty}D_z^\alpha \Phi(x,y,z) = \frac{1}{\Gamma(1-\alpha)} \frac{\partial}{\partial z} \int_{-\infty}^z \frac{{(x^2+y^2+u^2)}^{-1/2}\, du}{{(z-u)}^{\alpha}}\,\, .
\end{eqnarray}
Following the same steps in the previous subsection we conclude that
\begin{eqnarray}
_{-\infty}D_z^\alpha \Phi(x,y,z) = \frac{1}{\Gamma(1-\alpha)}\frac{\partial}{\partial z} \sum_{n=0}^\infty \frac{2n+1}{n(n+1) -\alpha(\alpha-1)} \frac{P_n(\cos\theta)}{r^{\alpha}}\, .
\end{eqnarray}
Writing ${\partial}/{\partial z}$ in terms of spherical coordinates
\begin{eqnarray}
\frac{\partial}{\partial z} =\cos\theta\frac{\partial}{\partial r} +\frac{\sin^2\theta}{r}\frac{\partial}{\partial \cos\theta}\,\, ,
\label{spherical}
\end{eqnarray}
we find
\begin{eqnarray}
 \frac{\partial^\alpha\Phi}{\partial z^\alpha} &=& \frac{1}{\Gamma(1-\alpha)\, r^{\alpha+1}}  \sum_{n=0}^\infty \frac{2n+1}{n(n+1)-\alpha(\alpha-1)} [ -\alpha\cos\theta P_n(\cos\theta) \nonumber \\
 &+&\sin^2\theta {P_n^\prime(\cos\theta)}] \, .
 \end{eqnarray}
Using the known identities of the Legendre polynomials [34]
\begin{eqnarray}
\sin^2\theta P_n^\prime =n\left( P_{n-1} -\cos\theta P_n\right)\, , \nonumber \\
(2n+1)\cos\theta P_n =(n+1) P_{n+1} +n P_{n-1}\,\, ,
\end{eqnarray}
and shifting the sum appropriately we reach the final result
\begin{eqnarray}
 \frac{\partial^\alpha\Phi}{\partial z^\alpha} = \frac{-\alpha}{\Gamma(1-\alpha) \,r^{\alpha+1}} \sum_{n=0}^\infty \frac{2n+1}{n(n+1)-\alpha(\alpha+1)} P_n(\cos\theta)\, .
\label{final1}
 \end{eqnarray}
The result is identical to the fractional integral, given in Eq.~(\ref{integral}). Thus the fractional integral and derivative of $\Phi=1/r$ have
the same form.
The result is valid for  $0 <\theta \leq \pi$ and $0\leq \alpha\leq 1$.
For $\alpha = 0$ it is easy to check that we retrieve the original field $1/r$, as all terms vanish except the first term (n=0). For $\alpha=1$ all terms vanish except the second term (n=1) and we retrieve the result  $-{P_1(\cos\theta)}/{r^2}=-z/r^3$, as expected.

Due to the spherical symmetry of the potential, one can easily conclude
\begin{eqnarray}
 \frac{\partial^\alpha\Phi}{\partial x^\alpha} =
 \frac{-\alpha}{\Gamma(1-\alpha)\, r^{\alpha+1}} \sum_{n=0}^\infty \frac{2n+1}{n(n+1)-\alpha(\alpha+1)} P_n(\sin\theta\cos\phi)
 \label{final2} \, , \\
 \frac{\partial^\alpha\Phi}{\partial y^\alpha} =
 \frac{-\alpha}{\Gamma(1-\alpha)\, r^{\alpha+1}}\sum_{n=0}^\infty \frac{2n+1}{n(n+1)-\alpha(\alpha+1)} P_n(\sin\theta\sin\phi)\, .
 \label{final3}
 \end{eqnarray}

The above results in Eqs.~(\ref{final1}, \ref{final2}, \ref{final3}) are valid for all values of $\alpha>0$.
Consider $m-1<\alpha<m$, where $m\in N$. Then
\begin{eqnarray}
 _{-\infty}D_z^\alpha \Phi  =
 \frac{1}{\Gamma(m-\alpha)}\int_{-\infty}^z \frac{{(x^2+y^2+u^2)}^{-1/2}\,du }{{(z-u)}^{\alpha+1-m}}
 \, \, .
\end{eqnarray}
Similar to the case of $0<\alpha<1$ we write $r^2 =x^2+y^2+z^2$ and let $t=z-u$ we get
\begin{eqnarray}
_{-\infty}D_z^\alpha \Phi =
\frac{1}{\Gamma(m-\alpha)}\int_0^{+\infty} \frac{{(r^2-2t\cos\theta+t^2)}^{-1/2}\,dt }{t^{\alpha+1-m}}
 \, \, .
\end{eqnarray}
Dividing the integration  into the two regions $0<t<r$ and $t>r$, we expand the integrand in terms of Legendre polynomials. Integrating over $t$
we get the final form
\begin{eqnarray}
_{-\infty}D_z^\alpha \Phi =  \frac{1}{\Gamma(m-\alpha)\, {r^{\alpha+1-m}}}\sum_{n=0}^\infty \frac{2n+1}{n(n+1) -(\alpha+1-m)(\alpha+m)} {P_n(\cos\theta)}\, .
\end{eqnarray}
Next we write $\partial^m/{\partial z^m}$ in terms of spherical coordinates, similar
to Eq.~(\ref{spherical})
\begin{eqnarray}
\frac{\partial^m}{\partial z^m} ={\left(\cos\theta\frac{\partial}{\partial r} +\frac{\sin^2\theta}{r}\frac{\partial}{\partial \cos\theta}\right)}^m
\label{spherical_2} \, .
\end{eqnarray}
 The calculation is tedious but for $ 1< \alpha<2$ we have explicitly performed the calculation and used few of the Legendre identities. We reached the same result in Eq.~(\ref{final1}), namely
 \begin{eqnarray}
 \frac{\partial^\alpha\Phi}{\partial z^\alpha} = \frac{-\alpha}{\Gamma(1-\alpha)\, r^{\alpha+1}} \sum_{n=0}^\infty \frac{2n+1}{n(n+1)-\alpha(\alpha+1)} P_n(\cos\theta)\, .
 \end{eqnarray}
For example for $\alpha=2$ all terms vanish except for $n=2$, thus it is straightforward to show that
${\partial^2\Phi}/{\partial z^2} = 2{P_2(\cos\theta)}/{r^3}$ as expected. In general one can show that
\begin{eqnarray}
\lim_{\alpha\rightarrow m} \frac{\partial^\alpha\Phi}{\partial z^\alpha}= (-1)^m \frac{m!}{r^{m+1}} P_m(\cos\theta)\, ,
\end{eqnarray}
as expected, where $m\in N$.

\section{Discussion and Conclusions}

We calculated the fractional integral and derivative of the potential $\Phi=1/r$. We found that for all values $-1< \alpha \leq 0$ and $\alpha\geq 0$,
\begin{eqnarray}
\frac{\partial^\alpha\Phi}{\partial z^\alpha} &=& \frac{-\alpha}{\Gamma(1-\alpha) \,r^{\alpha+1}} \sum_{n=0}^\infty \frac{2n+1}{n(n+1)-\alpha(\alpha+1)} P_n(\cos\theta)\, , \\
 \frac{\partial^\alpha\Phi}{\partial x^\alpha} &=&
 \frac{-\alpha}{\Gamma(1-\alpha)\, r^{\alpha+1}} \sum_{n=0}^\infty \frac{2n+1}{n(n+1)-\alpha(\alpha+1)} P_n(\sin\theta\cos\phi) \, , \\
 \frac{\partial^\alpha\Phi}{\partial y^\alpha} &=&
 \frac{-\alpha}{\Gamma(1-\alpha)\, r^{\alpha+1}}\sum_{n=0}^\infty \frac{2n+1}{n(n+1)-\alpha(\alpha+1)} P_n(\sin\theta\sin\phi)\, .
 \end{eqnarray}

One can implement
the fractional integral and derivative  of $1/r$ to relate two known mass distributions by a continuous
deformation, as discussed in Ref.~[27]. Given a mass distribution, $\rho(\vec{r})$, the
gravitational potential, $\Phi(\vec{r})$, can be determined by solving the Poisson equation
\begin{equation}
\nabla^2\Phi(\vec{r})=4\pi G \, \rho(\vec{r})\, ,
\label{eq1}
\end{equation}
where $G$ is the gravitational constant.
To illustrate this point we apply a fractional $\alpha$th-order differintegral operator,
with respect to the $z$ coordinate,
to both sides of Eq.~(\ref{eq1}). Taking the lower limit $a=-\infty$, we get
\begin{equation}
_{-\infty}D_z^\alpha\left[ \nabla^2\Phi(\vec{r})\right]=\nabla^2\left[
_{-\infty}D_z^\alpha\Phi(\vec{r})\right] =
4\pi G _{-\infty}D_z^\alpha\left[\rho(\vec{r})\right]\, .
\label{eq2}
\end{equation}
The commutativity of the two operators $\nabla^2$ and $_{-\infty}D_z^\alpha$ is maintained in our
case where the lower point is taken to be $a=-\infty$ and thus the potential $\Phi(\vec{r})$
and all its positive-integer derivatives at the lower point vanish. Thus the fractional potential $ _{-\infty}D_z^\alpha\Phi(\vec{r})$ corresponds to the fractional mass distribution $ _{-\infty}D_z^\alpha \rho(\vec{r})$.

Another application of the this work is to consider possible deviations of the inverse-square law gravitational field. Modifications are argued in theories of large extra dimensions, broken supersymmetry at low energy, and string theories (see Refs.~[28-30] and references therein). Consider the Newtonian gravitational potential $\Phi(r)=G m_1 m_2/r$. We consider the limiting case $\alpha=1-\epsilon$\, where $\epsilon<<1$ and choose the special case $\theta=\pi$, i.e. $r=z,\,x=y=0$.
Noting that $P_n(\cos\pi)={(-1)}^n$ and using the alternating harmonic series sum
\begin{eqnarray}
\sum_{n=2}^\infty {(-1)}^n \left(\frac{1}{1-n} -\frac{1}{2+n}\right) =\frac{-5}{6}\,\, ,
\end{eqnarray}
one can show that to a leading order of $\epsilon$
\begin{eqnarray}
 \frac{\partial^\alpha\Phi}{\partial z^\alpha} \approx \frac{G \,m_1 \,m_2 }{r^{2}}  \left(1+(\ln({\frac{r}{\lambda}})-1+\gamma)\epsilon\right)
 \end{eqnarray}
where $\gamma$ is the Euler number and $\lambda$ is a introduced for dimensionality.
Similarly, we can show that
\begin{eqnarray}
 \frac{\partial^\alpha\Phi}{\partial x^\alpha}= \frac{\partial^\alpha\Phi}{\partial y^\alpha} \approx \frac{G \,m_1 \,m_2 }{r^{2}} \, \epsilon
 \end{eqnarray}
The values of $\lambda$ and $\epsilon$ modify the Newtonian gravitational field and thus are restricted by existing experimental constraints.


\section*{References}

\begin{itemize}

\item[[1]]
  Fractional Integrals and Derivatives: Theory and Applications, S.G. Samko, A.A. Kilbas, O.I. Marichev, Gordon and Breach, New York (1993).

\item[[2]]
An Introduction to the Fractional Calculus and Fractional Differential Equations, K.S. Miller, B. Ross, John Wiley \& Sons, New York (1993).

\item[[3]]
Fractional Calculus: Integrations and Differentiations of Arbitrary Order, K. Nishimoto, University of New Haven Press, New Haven (1989).

\item[[4]]
The Fractional Calculus; Theory and Applications of Differentiation and Integration to Arbitrary Order, K B. Oldham, J. Spanier, Academic Press,  New York (1974).

\item[[5]]
 Fractional Differential Equations, I. Podlubny, Academic Press, San Diego (1999).

\item[[6]]
\label{Herrmann}
Fractional Calculus. An Introduction for Physicists, Richard Herrmann, World Scientific, Singapore  (2011).

\item[[7]]
 Applications of Fractional Calculus in Physics, R. Hilfer, World Scientific, Singapore (2000).

\item[[8]]
 Fractals and Fractional Calculus in Continuum Mechanics, A. Carpinteri, F. Mainardi (Eds.), Springer, Wien (1997).

 \item[[9]]
 Physics of Fractal Operators, B. West, M. Bologna, P. Grigolini, Springer, New York (2003).

 \item[[10]]
Review of Some Promising Fractional Physical Models, V. E. Tarasov, International Journal of Modern Physics B, Volume 27, Issue 09, 2013.

\item[[11]]
 Hamiltonian Chaos and Fractional Dynamics, G.M. Zaslavsky, Oxford University Press, Oxford (2005).

\item[[12]]
Multi�Dimensional Solutions of Space-Time-Fractional Diffusion Equations, A. Hanyga, Proc. R. Soc. Lond. 2002 vol. 458, 429-450.

\item[[13]]
The Differentiability in the Fractional Calculus, F. Ben Adda, Nonlinear Anal., 47 (2001).

\item[[14]]
Fractional Curl Operator in Electromagnetics, N. Engheta, Microwave and Optical Technology Letters, Volume 17, Issue 2, 1998.

\item[[15]]
Complex and Higher Order Fractional Curl Operator in Electromagnetics, Q.A. Naqvi, M. Abbas, Optics Communications, Vol. 241, 349-355, 2004.

\item[[16]]
Fractional Vector Calculus for Fractional Advection-Dispersion, M.M. Meerschaert, J. Mortensen, S.W. Wheatcraft, Physica A, 367, pp. 181�190, (2006).

\item[[17]]
Fractional Differential Forms, K. Cottrill-Shepherd, M. Naber, J. Math. Phys., 42, pp. 2203-2212, (2001).

\item[[18]]
Fractional Generalization of Gradient Systems, V.E. Tarasov, Letters in Mathematical Physics, Volume 73, Issue 1, pp 49-58 (2005).

\item[[19]]
Applications of Fractional Exterior Differential in Three-Dimensional Space, Yong Chen, Zhen-ya Yan, Hong-qing Zhang, Appl. Math. Mechanics, 24 (3) (2003), pp. 256-260.

\item[[20]]
Nonconservative Lagrangian and Hamiltonian Mechanics, F. Riewe, Physical Review E 53 (1996) 1890-1899 (1996).

\item[[21]]
Hamilton-Jacobi and Fractional Like Action With Time Scaling, M.A.E. Herzallah, I.M. Sami, D. Baleanu, and M.R. Eqab,
Nonlinear Dynamics 66 549-555 (2011).

\item[[22]]
On the Fractional Hamilton and Lagrange Mechanics, A.K. Golmankhaneh, M.Y. Ali, and D. Baleanu, International Journal of Theoretical Physics 51, (2012).

\item[[23]]
Hamiltonian Formulation of Systems with Linear Velocities Within Riemann�Liouville Fractional Derivatives, Sami I. Muslih,
D. Baleanu, Journal of Mathematical Analysis and Applications, Volume 304, Issue 2, (2005).

\item[[24]]
Lagrangian and Hamiltonian Fractional Sequential Mechanics, M. Klimek, Czech. J. Phys., 52 (2002).

\item[[25]]
Formulation of Euler-Lagrange Equations for Fractional Variational Problems, O.P. Agrawal, J. Math. Anal. Appl., 272, (2002).


\item[[26]]
On Fractional Calculus and Fractional Multipoles in Electromagnetism, N. Engheta, IEEE Transactions on Antennas and Propagation,
Volume 44,  Issue 4, (1996), 554-566.

\item[[27]]
Applications of Fractional Calculus to Gravity, Akram A Rousan, Ehab Malkawi, Eqab M Rabei, Hatem Widyan,
Fractional Calculus and Applied Analysis, Volume 5, Issue 2 (2002) 155-168.

\item[[28]]
Deviations from the $1/r^2$ Newton Law Due to Extra Dimensions, A. Kehagias, K. Sfetsos, Phys.Lett. B472 (2000) 39-44.

\item[[29]]
Sub-Millimeter Tests of the Gravitational Inverse-Square Law:
 A search for �Large� Extra Dimensions, C. D. Hoyle, U. Schmidt, B. R. Heckel, E. G. Adelberger, J. H. Gundlach, D. J. Kapner, and H. E. Swanson, Phys. Rev. Lett. 86, (2001) 1418.

\item[[30]]
Test of the Gravitational Inverse Square Law at Millimeter Ranges,Shan-Qing Yang, Bi-Fu Zhan, Qing-Lan Wang, Cheng-Gang Shao, Liang-Cheng Tu, Wen-Hai Tan, and Jun Luo, Phys. Rev. Lett., volume {108}, issue {8}, (2012) {081101}.

\item[[31]]
Axiomatic Basis for Spaces with Non-Integer Dimension, Frank H. Stillinger, Journal of Mathematical Physics, 1977.

\item[[32]]
Gravitational Potential in Fractional Space, Sami Muslih, Dumitru Baleanu, Eqab Rabei, Open Physics. Volume 5, Issue 3, (2007) 285�292.

\item[[33]]
Cantor-Type Cylindrical-Coordinate Method for Differential Equations with Local Fractional Derivatives,
Xiao-Jun Yanga, b, H.M. Srivastavac, Ji-Huan Hed, Dumitru Baleanue, Physics Letters A, Volume 377, Issues 28�30, 15 October 2013, Pages 1696�1700.

\item [[34]]
Mathematical Methods for Physicists, George B. Arfken, Hans J. Weber, Frank E. Harris,
Elsevier Academic Press, (2012).

\end{itemize}

\end{document}